\documentclass[prd,
               aps,
               floats,
               twocolumn,
               superscriptaddress,
               showpacs,
               nofootinbib,
               preprintnumbers] {revtex4}\input epsf
\usepackage{amsmath}
\usepackage{amsfonts}
\usepackage{graphicx}
\usepackage{dcolumn}
\def\be{\begin{equation}}
\def\ee{\end{equation}}
\def\ba{\begin{eqnarray}}
\def\ea{\end{eqnarray}}
\def\bs{\begin{subequations}}
\def\es{\end{subequations}}

\usepackage{color}

\begin{document}

\title{ Proposal for an experiment to search for
Randall-Sundrum type corrections to Newton's law of gravitation}

\author{Mofazzal Azam}
\affiliation{Theoretical Physics
Division, Bhabha Atomic Research Centre, Mumbai, India}
\author {M. Sami}
\affiliation{Centre
for Theoretical Physics, Jamia Millia Islamia, New Delhi, India}
\author{C. S. Unnikrishnan}
\affiliation{Tata Institute of Fundamental
Research, Mumbai, India}
\author{T. Shiromizu}
\affiliation{Department of
Physics, Tokyo Institute of Technology, Tokyo, Japan}

\begin{abstract}
String theory, as well as the string inspired brane-world models
such as the Randall-Sundrum (RS) one, suggest a modification of
Newton's law of gravitation at small distance scales. Search for
modifications of standard gravity is an active field of research in
this context. It is well known that short range corrections to
gravity would violate the Newton-Birkhoff theorem. Based on
calculations of RS type non-Newtonian forces for finite size
spherical bodies, we propose a torsion balance based experiment to
search for the effects of violation of this theorem valid in
Newtonian gravity as well as the general theory of relativity. We
explain the main principle behind the experiment and provide
detailed calculations suggesting optimum values of the parameters of
the experiment. The projected sensitivity is sufficient to probe the
Randall-Sundrum parameter up to $10$ microns.
\end{abstract}
\maketitle
Einstein's theory of gravitation is the theory of
space-time where the gravitational field is associated with the
space-time metric and curvature\cite{wein}. Although
phenomenologically an extremely successful theory, attempts to
quantize this geometric field have so far led to no decisive
progress. This difficulty has led many investigators to consider
higher dimensional theories in the hope that such attempts may help
to ultimately arrive at the quantum theory of gravitation in $(3+1)$
dimensions. String theory \cite{pol} and string inspired higher
dimensional theories such as the brane-world models \cite{ran} are
examples of such attempts. These theories suggest that the higher
dimensional effects would generally show up as a short range
correction to Newton's law of gravitation \cite{pol,ran}. Direct
astronomical observations and laboratory experiments had ruled out
corrections with range larger than a few millimeters even before the
recent surge of interest in higher dimensional theories. This leaves
possibility of corrections to Newton's law of gravity at millimeter
and submillimeter length scales \cite{fis}. Recent experiments are
steadily progressing to probe length scales down to $10$ microns.

In this paper, we will be concerned only with the $5$- dimensional
Randall-Sundrum (RS) brane-world model because it is  simple and
elegant, and it brings out the feature of the correction to
Newtonian gravity in a transparent manner \cite{ran,duff}. The RS
corrected potential is given by
\begin{eqnarray}
  U(r)= -\frac{Gm}{r}\left(1+\frac{l_{s}^{2}}{r^2}\right)
\end{eqnarray}
where the Randall-Sundrum parameter $~l_{s}^{2}=\frac{2}{3}l^2~$,
$~l$ is the curvature scale of 5-dimensional anti-deSitter
space-time, $G$ is Newton's constant of gravity, $m$ is mass and $r$
is the distance in 3-space. It turns out that these corrections do
not have any astrophysical significance(see Ref.\cite{azam} and
references there in). This leads us to conclude that, other than
accelerator based high energy experiments, direct observation of
this force in laboratories is the only way to test the presence of
such correction terms. We propose here a torsion balance based
experiment.

In the last two decades, several laboratory based experiments have
been carried out to verify the presence of corrections to Newtonian
gravity. The results in these experiments are generally
parameterized with an additional Yukawa interaction \cite{fis},
\begin{eqnarray}
  V(r)=-\frac{G m_1 m_2}{r}\Big{[}1 + \alpha~~exp(-\frac{r}{\lambda})\Big{]}
\end{eqnarray}
$\alpha$ being the strength of the additional interaction relative
to Newtonian gravity and $\lambda$ the range of the interaction.
These experiments set limit on the strength $\alpha$ for distance
scale $\lambda$, implying the absence of additional force whose
strength relative to Newtonian gravity, at distance scale $\lambda$,
is equal to or larger than $\alpha$. Even before the provocations
from string inspired models, in the years when a hypothetical 'fifth
force' was searched for, experimentalists had put stringent
constraints in the $\alpha - \lambda$ plane at length scale down to
a few $mm$ \cite{irvine,tifr,review1}. University of California at
Irvine group used "null-geometry" for torsion balance experiment and
set limit in the range : $\alpha=10^{-2}$ at $\lambda=3 mm$ to
$\alpha=10^{-4}$ at $\lambda=3 cm$ \cite{irvine}. More recently, the
Washington University group operated a specially designed "missing
mass" torsional pendulum experiment and set limit in the range:
$\alpha=10$ at $\lambda=100$ microns to $\alpha=10^{-2}$ at
$\lambda=3 mm$ \cite{wash}. "Cantilever" and "micro-cantilever"
based experiments have been carried out by the Colorado group and
the Stanford group respectively with constraints below $100$
microns\cite{canti}. There are also some experiments based on the
measurement of Casimir effect \cite{casimir}(see
Refs.\cite{review1,review2} for details).
\begin{figure}[h]
\centerline{\epsfxsize=17pc \epsfbox{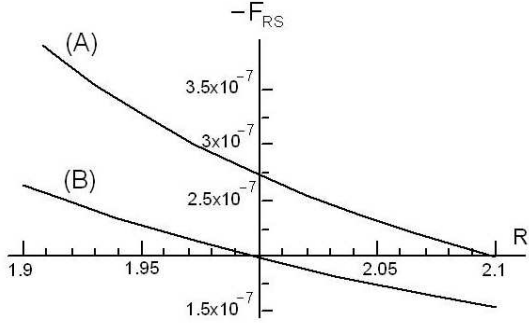} }
\caption{Force
between two spherical balls only due to the RS corrections as a
function of distance between their centres of masses. (A) Force
between 100 gm of silver ball and 10 gm of gold ball. (B) Force
between 100 gm of gold ball and 10 gm of gold ball. Force is given
in dyne and distance in cm . The RS parameter $l_s=1 mm~$ in both
cases.}
\end{figure}

The experiment we propose shares some features of the
"null-geometry" experiment of the University of California, Irvine
group. But we stress the importance of bulk spherical body in the
case of Randall-Sundrum gravity.
The main idea is that the Randall-Sundrum potential, like any other
short range potential, violates Newton-Birkhoff theorem. This theorem,
in our context, means that the effects of Newtonian gravity as well as
of general relativity of a spherically symmetric body depend only on
the mass and is independent of its size and the density of the
material \cite{wein}. With R-S potential, however, a spherically
symmetric body does not behave as a point source of gravity and
the potential as well force depends on density and size. Our
proposed experiment is intended to search for the quantitative and
qualitative outcome of violation of this theorem in the case of this
single parameter model.
We show that the short range corrections can lead to a measurable
effect for the numerical value of RS parameter $l_{s}$ at least up
to 10 microns.

In the following, we derive, in details, the Randall-Sundrum (RS)
interaction potential between two solid spheres of finite but
different radii and densities, separated by a distance, $R$, between
their centers. The RS potential $\phi_{RS}(r)$ of a spherical ball
of radius $a$ and constant density $\rho$, at a distance $r>a$ is
\begin{eqnarray}
  \phi_{RS}(r) &=& -Gl_{s}^{2}\int\frac{\rho(\vec{r'})d^3\vec{r'}}
                   {|\vec{r}-\vec{r'}|^3}  \nonumber \\
 &=& -2\pi G~l_{s}^{2}~\rho~\left[~ln\frac{r+a}{r-a}~-~\frac{2a}{r}\right]
\end{eqnarray}
which shows that the short range RS correction to gravity violates
Newton-Birkhoff theorem.\\
The force on a point mass $m$, is given by,
\begin{eqnarray}
  f_{RS} &=&-m\nabla\Phi_{RS}(r) =
  -3mGl_{s}^{2}\times\Big{[}\frac{M}{r^2(r^2-a^2)}\Big{]}\nonumber \\
 &=&-2\pi mGl_{s}^{2}~\times\frac{\rho}{\epsilon}
\end{eqnarray}
where the distance of the point mass $~r=a+\epsilon~$. Close to the
surface of the ball, the force is large and depends only on the
density of the source material. But away from the surface it falls
off very fast. Let us now consider two spherical balls of equal mass
$M$ but different radii $a_1$ and $a_2$, and densities $\rho_1$ and
$\rho_2$, $\rho_1>\rho_2$, $a_2>a_1$. Distance between the centers
of the two spheres is $2r$. A point mass $m$ is placed at the
midpoint on the line joining their centers. The Newtonian force of
spheres on the point mass test particles  balance each other. If
$f_1$ and $f_2$ are forces due to short range RS interaction, then
\begin{eqnarray}
\frac{f_2}{f_1}=\frac{r^2-a_{1}^{2}}{r^2-a_{2}^{2}}>1
\end{eqnarray}
\begin{figure}
\centerline{\epsfxsize=17pc \epsfbox{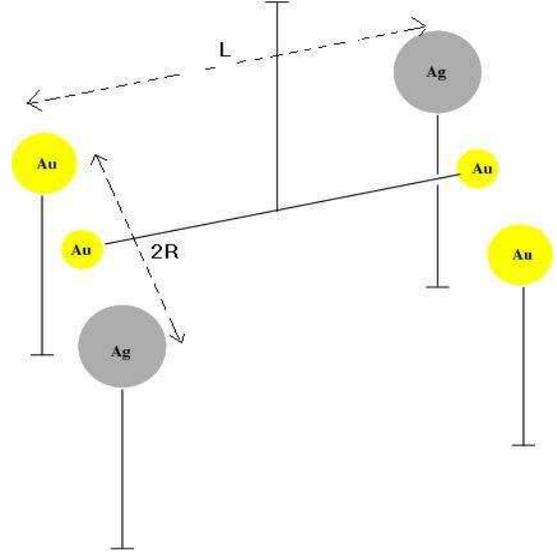} } \caption{The
experimental set up to measure the shift in the equilibrium position
of the torsion balance. The  ball at one end of the balance is
subjected to the combined force of Newtonian gravity and the RS
correction terms of the two balls symmetrically fixed at the same
end of the balance. Force on this ball due to the other two balls
fixed at the opposite end of the balance is negligible. $L= 20cm$.
$R$ varies with $l_s$. For $l_s=1~mm$, $R=2~cm$.}
\end{figure}
\begin{figure}
\centerline{\epsfxsize=16pc \epsfbox{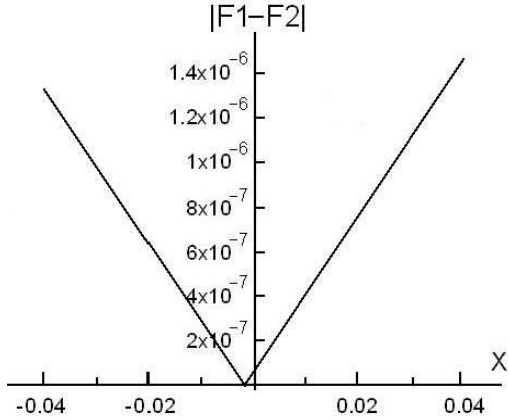} } \caption{Combined
force of Newtonian gravity and the RS-correction terms: The vertical
axis is the absolute value of the difference of combined forces (in
dynes) due to the fixed source masses, the $100~gm~$ gold ball and
the $100~gm$ silver ball on the $10~gm$ gold ball of the torsion
balance. The horizontal axis, $X$, is the distance (in cms) of the
centre of mass of the $10~gm$ gold ball from geometric midpoint
between the centres of the source masses. The RS parameter $l_s=1
mm~$. Details given in the text.}
\end{figure}
In the real experimental situations both the source mass and the
test mass have finite sizes. In what follows we shall calculate the
RS interaction potential between two spheres with radii $a$, $b$ and
densities $\rho_a$, $\rho_b$ respectively. Let the distance between
the centers of the spheres be $R$. Using Eq.(3), the RS correction
to the potential due to the two balls can be computed as
\begin{eqnarray}
&& \Phi_{RS}(R)=-2\pi~\rho_a~\rho_b~G~l_{s}^{2} \times\Bigg{\{}
\int_{0}^{b}~r^2~dr \int_{0}^{\pi} sin\theta d\theta \nonumber
\\
&&\int_{0}^{2\pi} d\phi \Bigg{(}ln\frac{|\vec{R}-\vec{r}|+a}
{|\vec{R}-\vec{r}|-a}
-\frac{2a}{|\vec{R}-\vec{r}|}\Bigg{)}\Bigg{\}}\nonumber
\end{eqnarray}
Integration over angles $\phi$, $\theta$,  and  radial parameter $r$
gives
\begin{eqnarray}
&&\Phi_{RS}(R)=-\frac{2\pi^2\rho_a\rho_bGl_{s}^{2}}{R}\Bigg{(}\Bigg{\{}
\Bigg{[}\frac{1}{4}(a^4+b^4)
-\frac{1}{2}a^2b^2 \nonumber\\
&&+\frac{1}{2}R^2(a^2+b^2)
-\frac{R^4}{12}\Bigg{]}~~ln\frac{R^2-(a+b)^2}{R^2-(a-b)^2}
~~\Bigg{\}}~~+\nonumber\\
&&\Bigg{\{}\frac{2R}{3}\Bigg{[}a^3~ln\frac{(R+b)^2-a^2}{(R-b)^2-a^2}
+ b^3~ln\frac{(R+a)^2-b^2}{(R-a)^2-b^2}\Bigg{]}\Bigg{\}} \nonumber\\
&&-\Bigg{\{}a^3b+\frac{1}{3}R^2ab+ab^3\Bigg{\}}~\Bigg{)}~~~~~~...~~~
...~~~...
\end{eqnarray}
The point mass test particle limit is obtained when $~b<<a~$ and
$~b<<R-a~$, $~M_b=\frac{4\pi}{3}b^3\rho_b~$, In this limit, the
expression above takes the form,
\begin{eqnarray}
&&\Phi_{RS}(R,a,b) =-2\pi
Gl_{s}^{2}\rho_aM_b\Bigg{(}ln\frac{R+a}{R-a} -\frac{2a}{R}\Big{[}1+
\nonumber \\
& &{\mathcal{O}}\Big{(}\frac{b^2}{R^2}(1-a/R)^{-4} \Big{)}
\Big{]}\Bigg{)}
\end{eqnarray}
\begin{figure}
\centerline{\epsfxsize=15pc \epsfbox{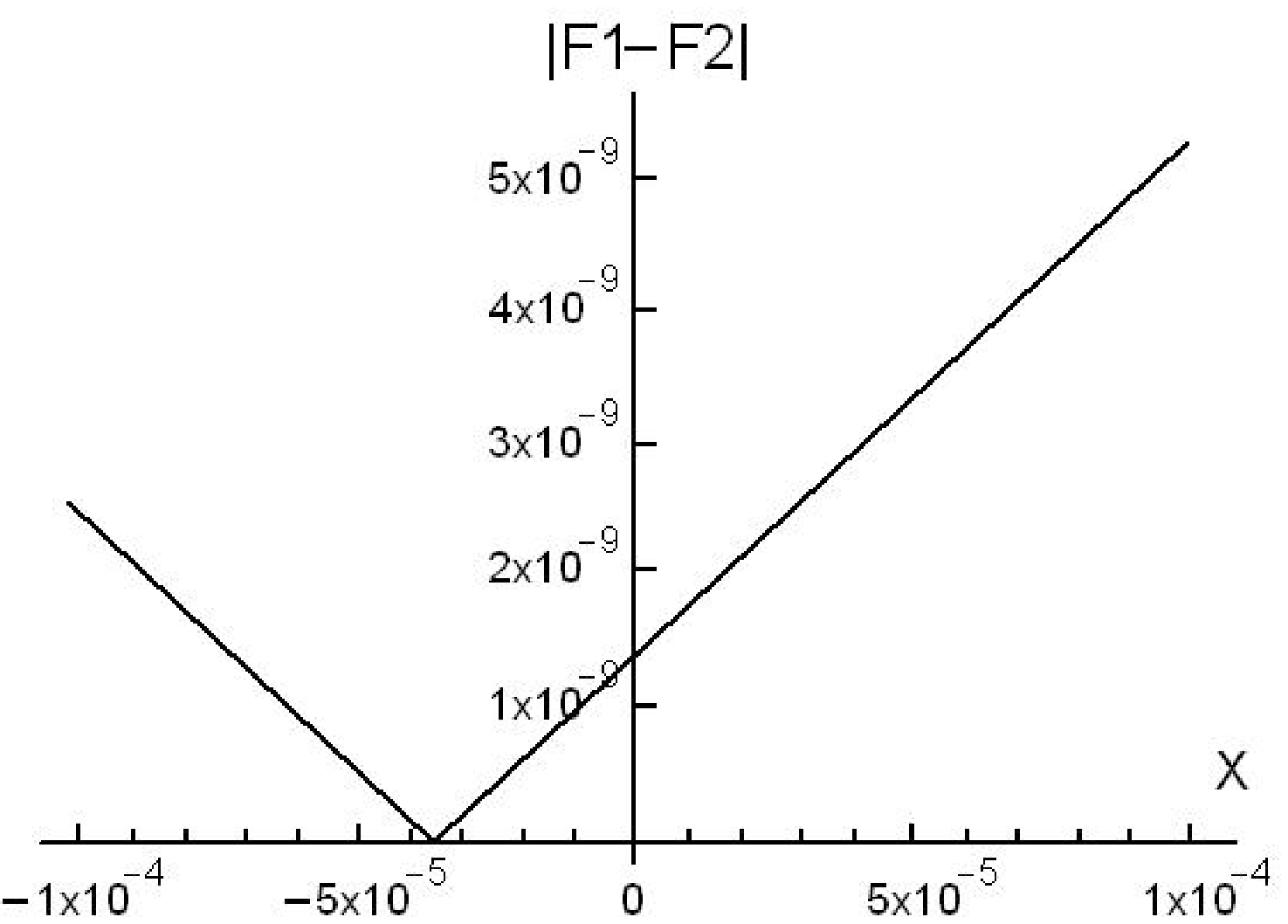} }
\caption{The plot is the same as the plot in Fig.3 with
RS parameter $l_s=100~microns~$. Details given in the text.}
\end{figure}
\begin{figure}
\centerline{\epsfxsize=15pc \epsfbox{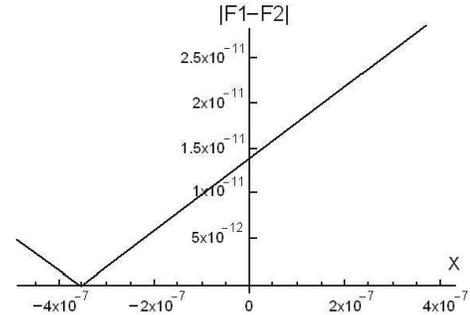} }
\caption{This plot is the same as the plot in Fig.3 with
RS parameter $l_s=10~microns~$. Details given in the text.}
\end{figure}
The potential given by Eq.(6) and the force generated by it
monotonically decrease to a finite value in the limit $R \rightarrow
a+b$. Thus the force as well the potential are finite when the balls
touch each other. In Fig.1 , we provide plots of the forces due to
the correction term. The vertical axis in the figure is $-F_{RS}$,
the absolute value of the force (in dynes), while the horizontal
axis is the distance between the centers of masses of the balls (in
cm). Plot (A) is the force, $-F_{RS}$, between 100 gm of silver ball
and 10 gm of gold ball, while plot (B) is the force, $-F_{RS}$,
between 100 gm gold ball and 10 gm gold ball. There is some
difference between the forces in the two cases considered. In
addition, the forces do not fall off too fast with the increase of
distance between the balls within the range favorable for a torsion
balance experiment. These are the features of RS corrections that we
want to exploit for our experiment. We emphasize that the magnitudes
of the forces and the relevant size scales are suitable for a
torsion balance experiment. An increase in density contrast of the
source materials or contrast in mass/density of the sources and of
the test body does not lead to any additional advantage.

A sketch of the scheme of the experiment is given in Fig.2. We have
four balls of $100~gm$ each placed in a planar rectangular
configuration in such a way that the centers of mass of the balls
are on the horizontal plane. The silver balls are diagonally
opposite to each other and so are the gold balls (radii of gold and
silver balls are $1.073$ cm and $1.315$ cm respectively). Along the
longer axis of the rectangle the distance, $L$, between the centers
of mass of the silver and gold balls is 20 cm, and along the shorter
axis the distance, $2R$, is 4 cm, a torsion balance hangs in the
middle, parallel to the longer axis of the rectangle. At each end of
the hanging bar of the torsion balance are attached two gold balls
of $10~ gm$ each with radius $0.498$ cm. The distance between the
centers of mass of these balls is 20 cm. The torsion coefficient of
the suspension wire can be taken to be about $~~0.1~dyne~cm/radian$.
In the absence of RS-correction term, the Newtonian force of the
$100~ gm$ silver and gold balls create unstable equilibrium point in
middle of the shorter axis of the rectangle at a distance of $R=2~
cm$ from either of the balls. In the presence of RS-correction, the
effect mentioned in the earlier paragraph would come into play and
the combined effect of Newtonian gravity and the RS-corrections
would shift the location of the unstable equilibrium point. Then the
torsion balance would oscillate about this shifted minimum of its
harmonic potential.

In Fig.3, we plot the absolute value of the difference of forces (in
dynes) due to the combined effect of Newtonian gravity and the
RS-correction terms of the fixed source masses, the $100~ gm$ gold
ball and  the $100~gm$ silver ball, on the $10~gm$ gold ball of the
torsion balance as a function of its distance (in cm) from the
geometric midpoint which is situated at a distance of $R=2~cm$ from
the centre of either of the source masses. It should be noted that
the unstable equilibrium position, where the combined force is zero,
moves by $20~ microns$ towards the $100~gm$ gold ball. In the
experimental configuration under consideration, this shift of the
equilibrium position towards the higher density ball is a
qualitative effect. Therefore, some systematic experimental
uncertainty can be eliminated by interchanging the positions of the
$100~gm$ gold and the silver balls. The equilibrium position should
again move towards the gold ball. The position of the unstable
equilibrium is found by locating the changed equilibrium position of
the torsion balance. The shift in the equilibrium position can be
increased by decreasing the distance between the fixed $100~ gm$
gold and the $100~ gm$ silver balls along the shorter axis
rectangular configuration but leaving the distance along the longer
axis unchanged. For example, a distance of 3.8 cm with midpoint at
1.9 cm, the equilibrium position shifts by 35 microns. A further
decrease can give a shift of about 50 microns, after which the
atomic forces start to interfere.

The accuracy of the angular shift that can be measured with standard
technology in a torsion balance experiment is below
$~10^{-9}~~rad/\sqrt{Hz}~$ which for our configuration amounts to a
distance shift of the end point of the balance of about
$~10^{-8}~~cm$ which is several order of magnitude smaller than the
shift in the case when RS parameter $l_s=0.1~cm$. Systematic effects due to Newtonian
gravity gradients arising from errors of about 5 microns in the position of the source masses, and due to the deviations from their sphericity and density homogeneity at the level of $10^{-3}$ generate less than $100~ nm$ shift in the equilibrium position of the test mass \cite{chen}. The small drift of
the torsion balance, amounting to $1~microradian$ per hour, can also
be corrected at this level in repeated measurements \cite{dicke}.
Therefore, achieving required sensitivity to detect RS corrections
for $l_{s}=100~microns$ is not difficult. To probe RS corrections
for $l_{s}=10~microns$ the masses have to be located accurate to
less than $1~micron$ and the drift should be corrected at 1$\%$
level, which is feasible but requires considerable care in
experimental design. This sets the  lower limit on the value of RS
parameter that can be probed with some reasonable degree of
confidence to about $l_{s}=10~microns$. This can be inferred from
figures Figs.4 $\&$ 5. These figures correspond to the case when the
length of shorter rectangular axis in the setup in Fig.2 is
$3.8~cm$.

 We have discussed in this paper the basic
principle, feasibility and the schematic of the experiment. A torsion balance experiment
along the lines discussed in this paper is under active
consideration at the Tata Institute of Fundamental research, Mumbai.

 We thank M. J. Duff, K. Kuroda, N. Mitskievich and H. Tagoshi for
their useful comments. MS and TS are supported by DST/JSPS Grant No
DST/INT/JSPS/Project-35/2007.

\end{document}